# SLOTS-MEMENTO: A SYSTEM FACILITATING INTERGENERATIONAL STORY SHARING AND PRESERVATION OF FAMILY MEMENTOS

Cun Li, Jun Hu, Bart Hengeveld, Caroline Hummels

Department of Industrial Design, Eindhoven University of Technology, Eindhoven,Netherlands

## ABSTRACT

*Family mementos document events shaping family life, telling a story within and between family members. The elderly collected some mementos for children, but never recorded stories related to those objects. In this paper, in order to understand the status quo of memento storytelling and sharing of elderly people, contextual inquiry was conducted, which further helped us to identify design opportunities and requirements. Resulting design was defined after brainstorm and user consultation, which was Slots- Memento, a system consisting a slot machine-like device used by the elderly and a flash drive used by the young. The Slots machine-like device utilizes with the metaphor of slots machine, which integrates functions of memento photo displaying, story recording, and preservation. In the flash disk, the young could copy memento photos to it. The system aims to facilitate memento story sharing and preservation within family members. Preliminary evaluation and user test were conducted in evaluation section, the results showed that Slots-Memento was understood and accepted by the elderly users. Photos of mementos were easy to recall memories. It enabled the elderly people to be aware of the stories of the family mementos, as well as aroused their desire to share them with family members. Related research methodology includes contextual inquiry, brainstorming, prototyping, scenario creation, and user test.*

## 1. INTRODUCTION

A memento is an object given or deliberately kept as a reminder of a person, place or event and it is directly meaningful to their memories [1]. When the same object was mentioned by young and old, young people connected to it actively, older people contemplatively [2]. Family mementos are highly valued, and support different types of family stories and recollections [3].

Storytelling of family mementos serves multiple functions to the elderly. Firstly, from the perspective of memento itself, mementos have strong associations with past ev, people or habits, and they help people reflect through vivid re-experience of their past [4]. In addition, recalling memories of mementos is a process of reminiscence, which improves psychological well-being and helps older adults find meaning in their life [5]. Secondly, from a social perspective, stories told by the elderly create meaning beyond the individual and provide a sense of self through historical time and in relation to family members, and thus may facilitate positive identity [3]. Thirdly, from a broader perspective, mementos and related stories are important part of identity preservation. Dying people hope they will be remembered, however when death occurs, survivors are left with bundles of images, materials, objects, and wishes of the deceased [6].

There is an increasing prevalence of personal digital content, Facebook, Twitter, etc., along with the growing tendency in human–computer interaction (HCI) toward designing technologies for homes, families, and experiences [5]. However, these platforms are more about the “now” moments and less about the past moments [7]. In addition, the elderly are still disconnected from the mainstream social circles due to the lack of technology and devices that resonate with them [8].

In this paper, we present Slots-Memento, a system consisting a slot machine-like device used by the elderly and a flash drive used by the young. The system aims to facilitate memento story sharing and preservation within family members. The Slots machine-like device utilizes with the metaphor of Slots machine, which integrates functions of memento photo displaying, story recording, and preservation. In the flash disk, the young could copy memento photos to it. It is also used to preserve story audios. It could be used by the elderly and their family members either together or separately.

## 2. RELATED WORK

### 2.1 MEMENTOS AND RELATED MEMORIES

Related work of mementos and related memories could be concluded from both theoretical and practical aspects.

On the theoretical side, a fieldwork where participants gave us a tour of their homes describing how and why particular objects become mementos[1]. An empirical study of parents’ ability to retrieve photos related to salient family events from more than a year ago[9]. A fieldwork to characterise and compare physical and digital mementos in the home[3]. discussing results from fieldwork with different community groups in the course of which seemingly any object could form the basis of a meaningful story and act as entry point into rich inherent ‘networks of meaning’[10].

On the application side, related research and application could be classified into three aspects. Firstly, there are lifelogging technologies which capture and archive personal mementos and memories: A study using SenseCam, a “life- logging” technology in the form of a wearable camera, which aims to capture data about everyday life in order to support people’s memory for past, personal events [11]; Secondly, organizing and archiving digital representations of mementos relating to people, places and objects: MemoryLane, which allows people to capture, actively organize and reflect on digital representations of mementos relating to people, places and objects [12]. Thirdly, making connections between objects and stories based on barcodes and RFID technology: Living Memory Box, a device and service to assist families in preserving memories in a variety of media forms[13].

### 2.2 ELDERLY AND TECHNOLOGY

According to the literature, reasons the elders do not use digital devices could be concluded: anxious about using technology, age-related physical decline, and user-unfriendliness of technology.

#### 2.2.1 ANXIOUS ABOUT USING TECHNOLOGY

Anxious about using technology. Despite their willingness to learn how to use a computer, most older people still regard technology as something not belonging to their own world, feeling uncomfortable and anxious about it [14].

#### 2.2.2 AGE-RELATED PHSICAL DECLINE

Due to the limitations resulting from the psychological and physiological process of aging, the elderly users face many problems when interacting with devices [15]. In addition to attitudinal variables, cognitive abilities are important to technology adoption for the elderly users [16].

#### 2.2.3 USER-UNFRIENDLINESS OF TECHNOLOGY

Currently most interfaces are designed to support younger users. To support elderly users, more age-related differences are needed to be considered.

### 2.3 METAPHOR IN INTERFACE

Research shows that elderly users prefer using technologies that are recognizable in their everyday lives. A digital bulletin board was offered to the elderly users in the experiment conducted by Erwin and Veldhoven. The experiment indicates that the familiar metaphor design could reduce anoxia about the usage of new product in elderly subjects as well as reduce learning time in elderly subjects [17].

### 2.4 SUMMARY

Firstly, there is few work and design that focus on mementos of the elderly. Little attention has been paid to excavating stories behind mementos with an eye on designing a system to facilitate memento story sharing and preservation within family members. Secondly, literature of elderly and technology implies that elderly people are a special group that more requirements need to be considered in design. Thirdly, we could draw inspirations from the literature of interactive installation that metaphor is an effective way to lower thresholds of technology access

## 3. CONTEXTUAL INQUIRY

### 3.1 SEMI-STRUCTURED INTERVIEW

To understand the status quo of memento storytelling and sharing of elderly people, define design requirements, and further identify design opportunities, the method of contextual inquiry was adopted. Contextual inquiry is a semi-structured interview method to obtain information about the context of use, target people are interviewed in their own environments, the analysis data is more realistic than laboratory data [18].

We conducted 11 semi-structured interviews with the older residents in a local nursing home and each interview was audio-recorded after obtained the interviewees' approval. They ranged in age from 65 to 86 years, as is shown in Figure1. The following topics were discussed: Basic information, communication with family, Current memento story sharing situation, and familiarity with technology, detailed questions are shown in Table 1.

Table 1. Interview topics in contextual inquiry.

| |
|---|
| **Basic Information** |
| Age, gender, physical condition |
| **Communication with family** |
| Who, number, frequency, duration of contacting with family members |
| Way of keeping in touch (face-to-face, phone, skype,.etc) |
| **Current memento story sharing situation** |
| what mementos do you still keep and why? |
| Have you ever told stories behind the mementos to others? Why? |
| In which situations, for what reasons you will share the memento stories? |
| who, when, where, and how to talk (face to face ? telephone ? Skype)? |

## 3.2 FINDINGS IN CONTEXTUAL INQUIRY

### 3.2.1 COMMUNICATION WITH FAMILY

With the exception of 1 interviewee has no children, all of them have more than 1 child. They have regular contact with children: their children visited them weekly and duration of every visit was 3-5 hours. Their connections with friends were less as their friends, brothers and sisters were also aged or passed away, connections between them were less: *"I have one sister, my younger sister， contact with her less and less because they are dying. Also, I am old and couldn't take train or drive car."* They also connected through telephone, but they called their children only when under special or urgent circumstances (diseases, holidays). One elderly said: *"I don't call my children very often because I don't want to disturb them, they have to work after all."*

### 3.2.2 CURRENT MEMENTO STORY SHARING SITUATION

They would like to share mementos with others. When asked If they would like to share mementos and albums to others, one interviewee said: *"Yes I do. Luckily I was able to save my photo albums. I also had a book with pictures from, I used to do dancing competitions, but I lost it, it's a shame."* One interviewee was not native and had an album of hometown photos: Another interviewee said: "I am interested in photos of my hometown, and I have three picture books of my hometown. During the historical afternoon, we shared hometown pictures with others."

### 3.2.3 PROBLEMS ENCOUNTERING WHEN SHARING MEMENTOS

Most older residents kept albums and small mementos in nursing home. However, most of their family mementos weren't brought to nursing home when they moved here

### 3.2.4. FAMILIARITY WITH TECHNOLOGY

Most elderly interviewees couldn't operate computers or smartphones. The major causes were that the elderly were unfamiliar with them and lack of using experiences. The major causes were that the elderly were unfamiliar with them and lack of using experiences. They obtained information mainly by TV and Newspapers. Elderly people still relied heavily on paper and prefer physical interaction and operation.

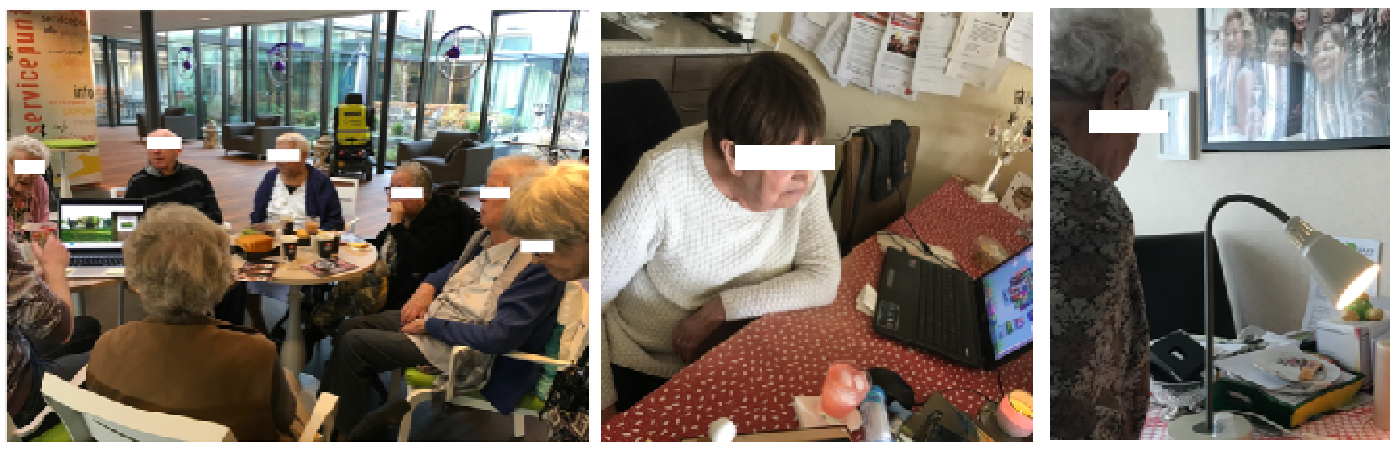

Figure 1. Semi-structured interview with the elderly.

### 3.3 Design Requirements

Our target group is the elderly, who have some certain specificities compared with the young generation. Based on the above literature review, the following design requirements need to be taken into accounts.

#### 3.3.1 Tangible Interface

According to our preliminary research, most elderly interviewees couldn't operate computers or smartphones. The major causes are, on the one hand, the elderly are unfamiliar with them and lack of using experiences. On the hand, touch-based interfaces are mainly visually guided without physical feedback. This lack of tangible feedback leads often to several errors and frustrations that are accentuated in older people [19].

#### 3.3.2 Using Metaphor In Interface

Elderly people are greatly interested in traditional physical objects such as handicrafts and radios. A new product that is designed in familiar metaphors can reduce the barriers of elderly users to use it[20] . To the novice users, gestural interaction provides for easier and more enjoyable learning and remembering[21].

#### 3.3.3 Intuitive Interaction

There exists a link between intuition and experience. The interaction style could be based on the elderly's familiar knowledge to help them understand easily. As the nostalgic feelings are pervasive in the elderly people, the metaphor could be adopted to enhance familiarity and simplicity [22].

## 4.IMPLEMENTATION

### 4.1CONCEPT

The capacity of storytelling depends on two main factors: narrative competence and memory capacity [23]. In the concept of the prototype, a simple operation of story recording is applied to lower the cost of narrative of the elderly, metaphor is applied in interface to reduce learning time, and gesture operation is applied to make the interaction intuitive. As is shown in Figure 2.

## 4.2 BRAINSTORM

Brainstorm method was used to generate ideas as many as possible. All the design ideas should be based on the design concept as well as meet the previous design requirements. Three concepts were chosen after an initial selection round.

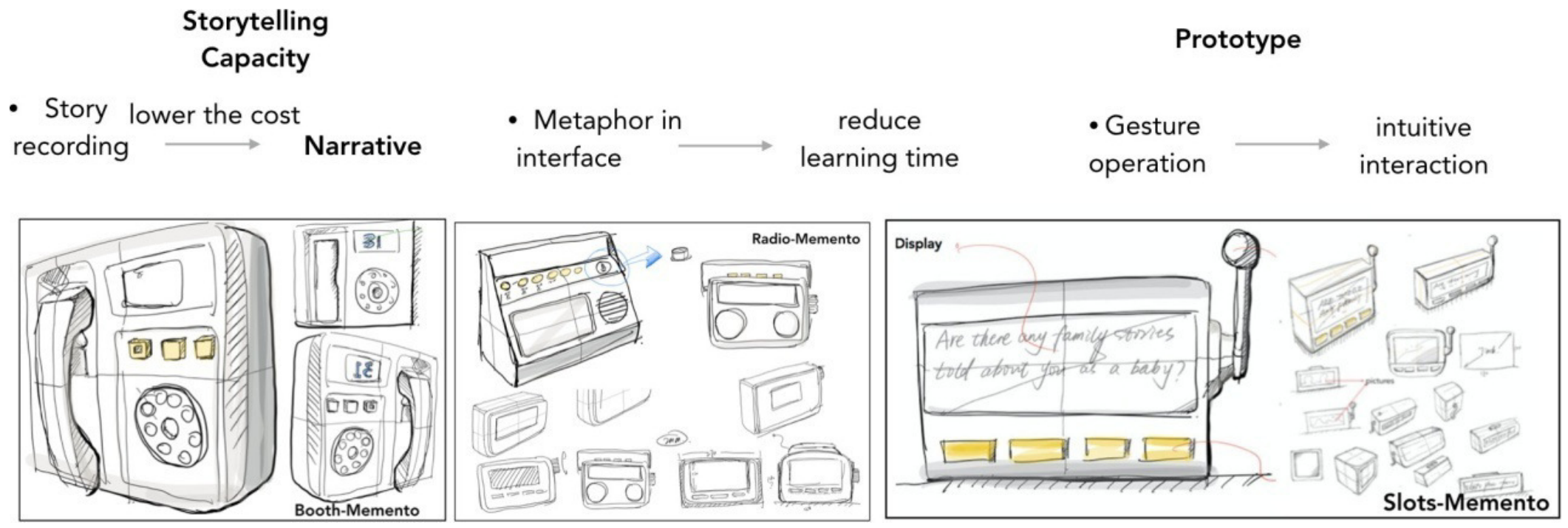

Figure 2. Proposal sketches in brainstorm phase.

### 4.2.1 BOOTH-MEMENTO

It is an installation utilized with the metaphor of antique cradle telephone. Pick up the receiver and turn the dial, a memento photo will be displayed. The elderly users tell stories related to the memento and the stories will be recorded.

### 4.2.2 RADIO-MEMENTO

It is an installation utilized with the metaphor of antique radio. Memento photos display on the screen of it. Elderly users could switch memento photos by turning the knob. Stories told by the elderly will be recorded after pressing the recording button.

### 4.2.3 SLOTS-MEMENTO

It is an installation utilized with the metaphor of slot machine. Memento photos displays when user pulls down the handle. Stories told by the elderly will be recorded after pressing the recording button.

## 4.4 USER CONSULTATION

To evaluate the above design ideas and get feedback, elderly people from the local nursing home were consulted. Design ideas were introduced to the interviewees in details. To make the elderly better understand our design ideas, the above sketches and related pictures including antique cradle telephone, antique radio and slots machine were showed to them as well.

We expected to know the following information in this interview: familiarity of the three objects, operations, preferences and suggestions for improvement.

All the interviewees were familiar with the above objects. When talking about the ease of operation, one interviewee said: *I used to listen to radios, and I still have one in my room. But I feel difficult to operate precisely when I scroll through stations by turning a knob.* When asked about the preference of the three ideas, one interviewee said: *I prefer the Slots-* *machine because the operation is interesting.* In brief, most elderly people reflected that the handle operation is more intuitive.

In addition to the opinions of the elderly, the above three design concepts are assessed also from the following aspects: innovation, ease of operation and feasibility. Based on the interview feedback and the assessment, concept Slots-Memento was chosen and further detailed.

### 4.5.SLOTS-MEMENTO DEVICE USED BY THE ELDERLY

Research shows that Familiar metaphor design could reduce anoxia about usage of new product as well as reduce learning time in elderly subjects [17]. Slot machine is operated by one lever on the side of the machine, which was invented in 1891 and is familiar to most elderly people. Metaphor of slot machine is adopted in the design of Slots-Memento, which is a novel device that borrows from the way of operation of slots machine, which is easily accessible and use for them. Just like the slot machine, a memento photo displays when the user pulls down the handle in the operation of Slots-Memento, which provides what the elderly users understand the way of interacting, rather than introduce a new interface to them. The gestural interaction in Slots- Memento offer easier and more enjoyable learning and remembering [21].

#### 4.5.1 APPEARANCE

The prototype is L-shaped with rounded corners, portable dimension enables it to move conveniently (Figure 3). Moreover, built-in power makes it do not require external infrastructure to operate. There is a 7-inch display on the front and lever with knob is on the right. From left to right in order, power button, microphone, and “REC/STOP” button are arranged below the screen.

The shell is made of porcelain white acrylic. To make it with vintage and decorative effect, external surface of it is covered with wood-grained paper, which is also in line with the aesthetic view of the elderly.

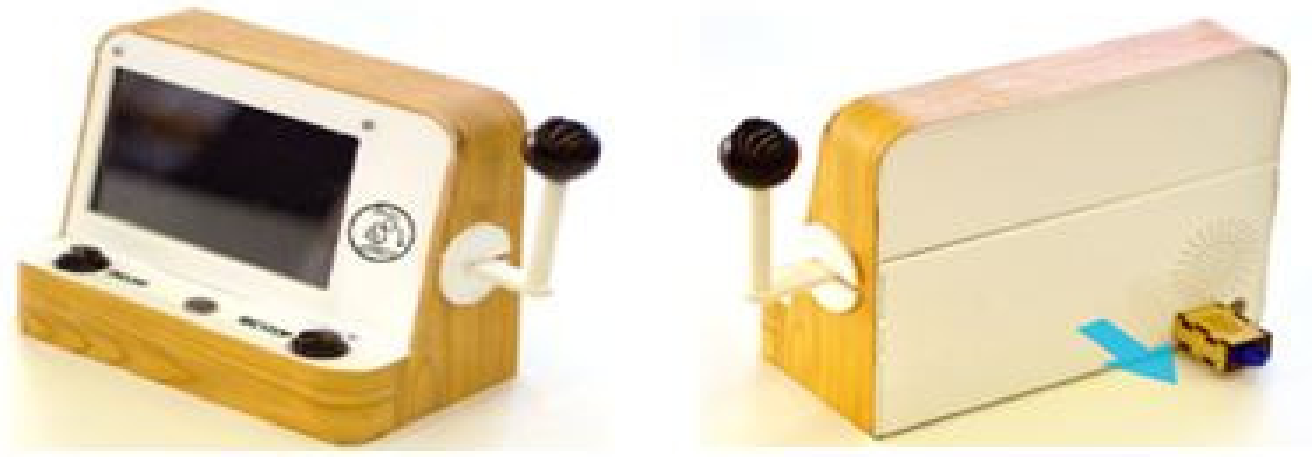

Figure 3. Prototype of Slots-Memento

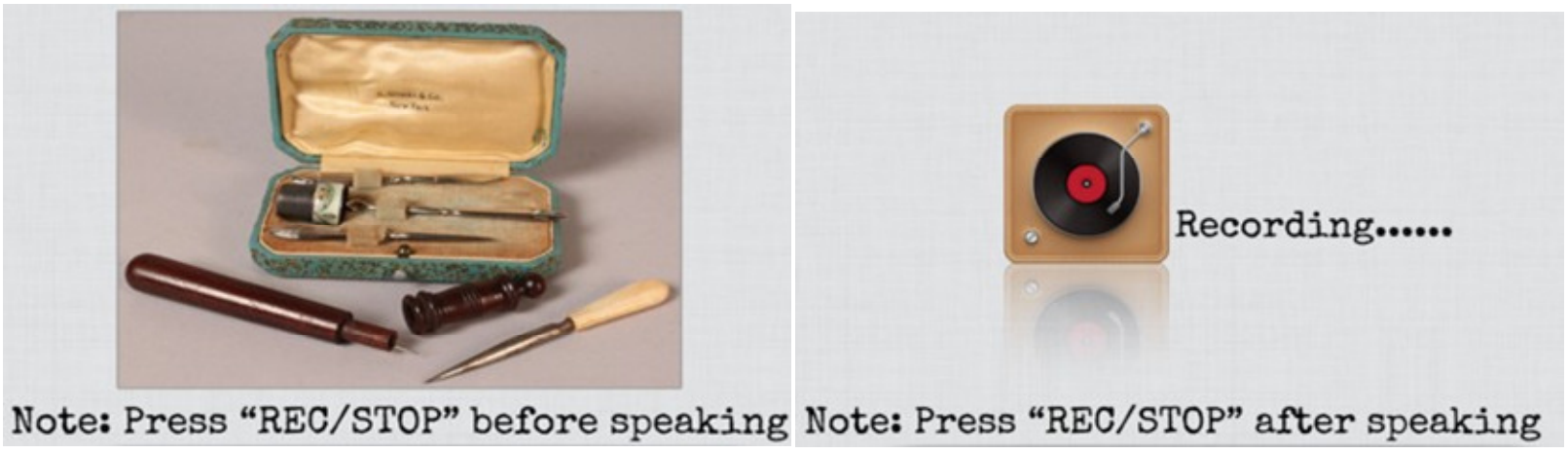

Figure 4. Graphical interfaces of Slots-Memento.

#### 4.5.2 GRAPHICAL INTERFACES

There are two graphical interfaces in Slots-Memento: the “Photo interface” and “Recoding interface” (Figure 4). Vintage style is also applied in both in the interface elements and fonts to make them integrated into the design style of Slots-Memento device. In addition, it is rendered in huge and bold fonts considering the fading eyesight of the elderly. There are usage tips at the bottom: “Note: Press “REC/STOP button before/after recording”

The “Photo interface” displays one specific memento. It would be switched to next/previous photo by pulling down/pushing up the knob. The “Photo interface” would be switched to “Recoding interface” if the REC/STOP button are pressed. In the “Recoding interface”, a dynamic recording icon and timer widget are placed to provide real-time feedback.

#### 4.5.3 OPERATING PROCEDURES

Interaction is strongly associated with content organization. Literature shows that performance of retrieving digital pictures was relatively poor for parents[9]. Photo organization in Slots-Memento is based on the metaphor of slot machine, which enables the photo navigation intuitive and easier: the elderly user is able to explicitly browse a picture by pulling up/down the handle.

To make the interaction simple and intuitive, there are only two operation components: a lever and a big button. As is shown in Figure 5, the detailed operating process is: (1) The young copy photos of family mementos to flash drive, and insert flash drive into Slots-Memento device. (2) The elderly press the left button to turn it on. (3) The elderly pull down/push up the knob to switch to the next/previous photo. (4) The elderly press the “REC/STOP” button to start recording stories related to the photos. (5) The elderly press the “REC/STOP” button again to save the recordings. Story audios are saved in the Flash-drive. (6) The young plug the flash disk into a computer to listen and keep stories, and further add other photos of mementos.

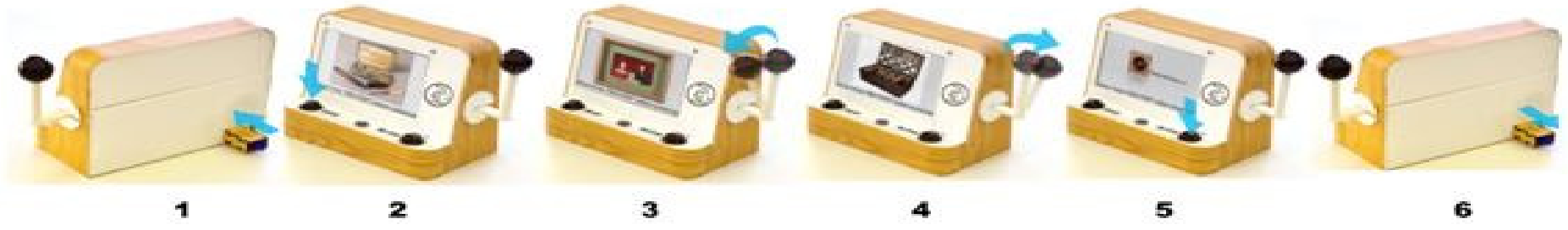

Figure 5. Operating procedures of Slots-Memento

#### 4.5.4 HARDWARE OF SLOTS-MEMENTO

Raspberry Pi 2 Model B is chosen as the hardware platform for Slots-Memento (Figure6). Raspberry Pi provides core processing functionalities; Joystick USB Encoder board is the medium to connect Raspberry Pi and joystick. Assembly of microphone and sound-card provides audio input. Slots-Memento is powered by Power Bank. Lever is 3D printed which could fit into the joystick component. The LCD Screen is graphical output.

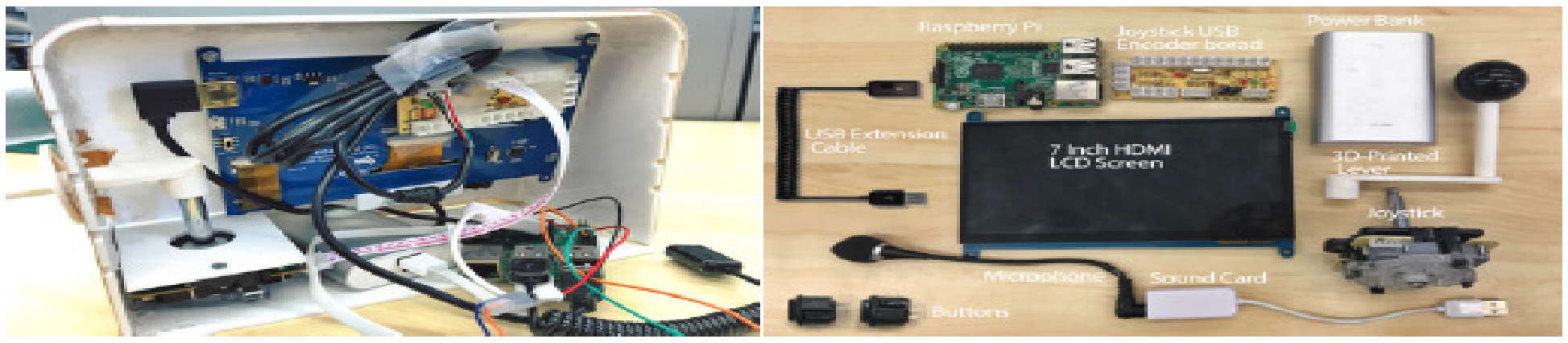

Figure 6. Internal structure and structure of Slots-Memento.

## 4.6 Flash Drive Used By The Young

Flash drive is embedded in a MDF-made shell (Figure 7). The flash disk is used to store memento photos and preserve story audios. There are two folders in the USB Flash Disk: “MEMENTOS” and “RECORDINGS”, the former contains the photos of mementos, the latter contains all the story audios told by the elderly.

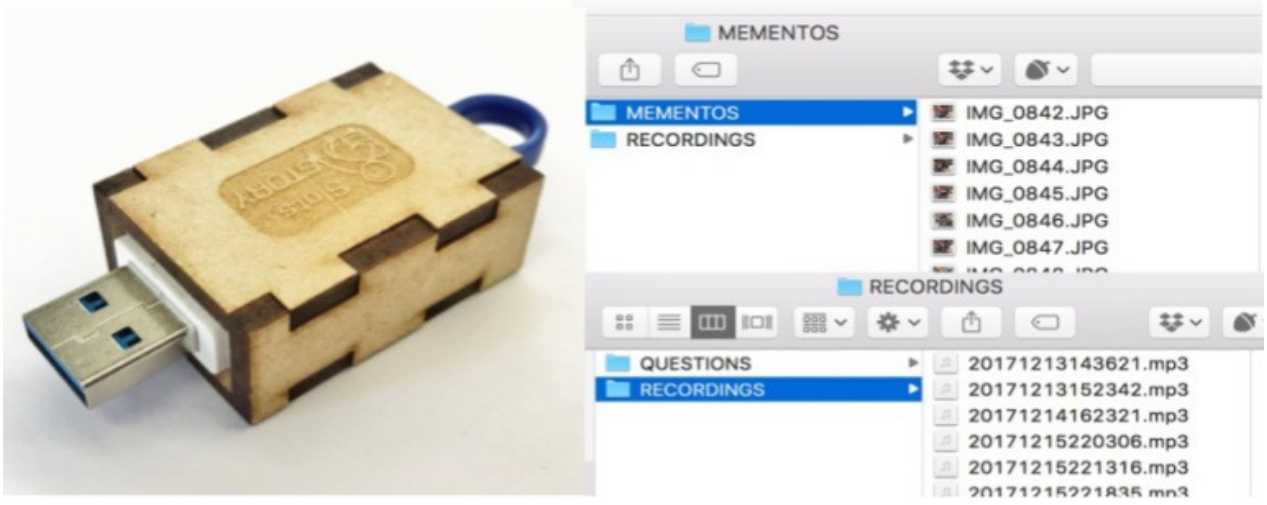

Figure 7. USB flash disk and folders in it.

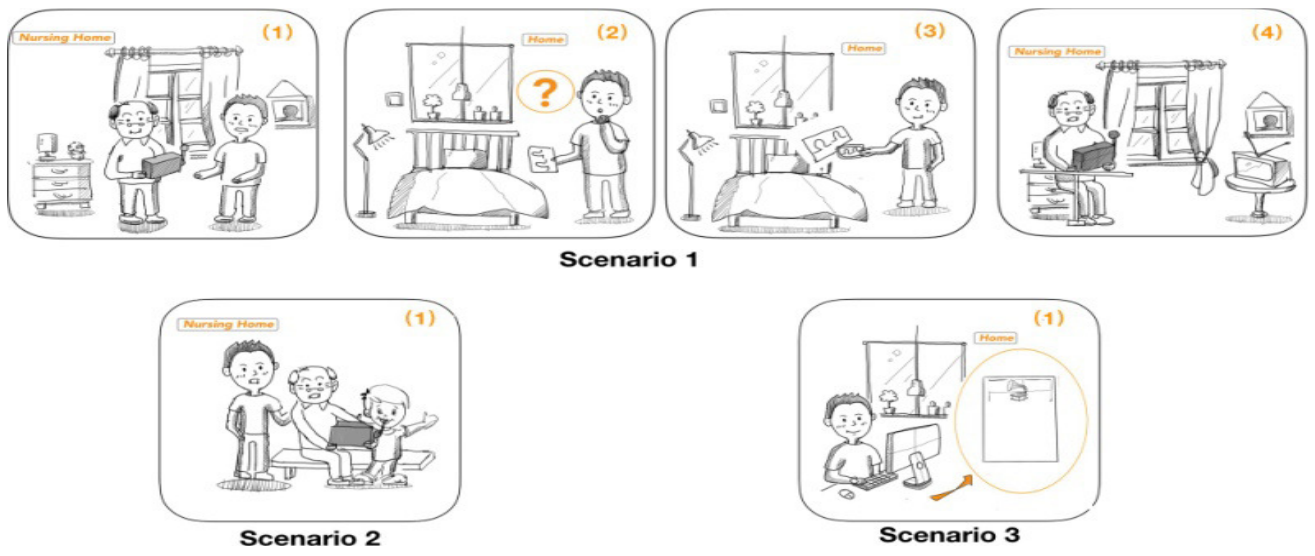

Figure 8. Storyboard of Slots-Mement

## 4.7 Usage Scenario Of The Slots-Memento System

David grows old and lives alone. Someday his son gives him Slots-Memento and he soon masters functions of it after his son’s short instruction (Figure 8).

### 4.7.1 Share Memento Stories Over A Distance

David chooses mementos photos by pulling down and pushing up the lever. He tells the stories behind the family mementos and all the stories are saved in the flash disk. David’s son gets the flash disk then and plugs it into a computer to listen the recordings. With great interests, his son wants to know more details of a photo of David. He picks up the phone and calls his David: “Hey Dad, I listened to your story of the photo of you, I am very interested and I want to know more about it.” David is then told more details of that story and his son feels he knows more about his father.

#### 4.7.2 Share Mementos Stories Face To Face

Every weekend David’s son and his grandson go to visit David, which is the happiest moment for Him. David’s grandson likes him because he could always hear lots of stories from David. With the mementos provided by Slots-Memento, the whole family get together and listen to stories of their family mementos.

#### 4.7.3 Modify Memento Photos

David has been using Slots-Memento for a long while and almost all the stories of mementos are told. His son then plugs the flash disk into a computer to preserve stories and add more memento photos, and then give the Slots-Memento back to his father.

## 5. Evaluation

### 5.1 Preliminary Evaluation

In preliminary evaluation phase, discovery of ideas and insights is in priority, rather than collecting statistically accurate data. To be specific, firstly, purpose, functions and operation procedure were introduced to elderly in a local nursing home. Secondly, the prototype was then operated by them. Thirdly, short interviews were conducted with the elderly and their family members. Our main objective of the interview is to obtain information from the following aspects: Validity, contents, appearance, interaction, comments and open questions of Slots-Memento. Detailed interview topics are shown in Table2.

| |
|---|
| **Validity** |
| Would you like to use it? who would you like to share with? |
| Do you think it could facilitate sharing stories of mementos? why? |
| **Appearance** |
| What kind of appearance style would you prefer:<br>A. Vintage B. High-tech C. Colorful/Lovely D. Simple |
| **Interaction** |
| Do you understand the concept of the prototype ? |
| Do you find it easy to use? what is the most difficult part? |
| **Comments** |
| Which part do you like / dislike most of the prototype? why? |
| Which parts would you change the prototype? Why? |
| **Open questions** |
| Would you like to use it face to face/over a distance? |
| Do you want to share mementos/photos with your children? |

Table 2. Interview topics in evaluation.

### 5.2 User Test

User test was then designed and conducted. Two family was involved in the user test. the following mementos were collected: handicraft, photograph, postcard, note, booklet, map, and document. Short interviews were also conducted with the young people afterwards.

## 5.3 RESULTS

### 5.3.1 MEMENTOS COLLECTED FROM THE INTERVIEWEES

The following kinds of mementos were collected in the user test phase: handicraft, photograph, postcard, note, booklet, map, and document. Photographs were the most common mementos of the elderly people. The second were handicrafts they collected or made by themselves (Table 3).

Table 3. Mementos collected from the interviewees.

| Handicrafts | Photograph | Postcard | Note | Booklet | Map | Document |
|---|---|---|---|---|---|---|
| 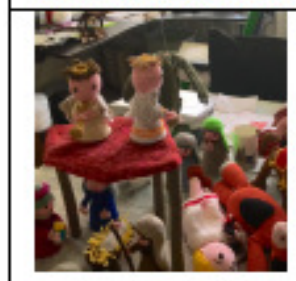 | 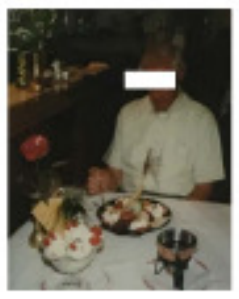 | 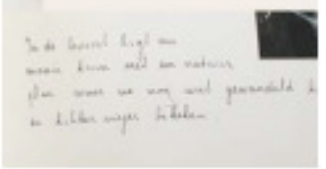 | 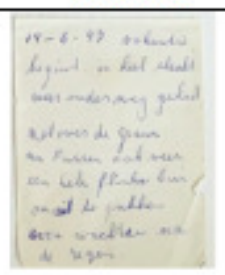 | 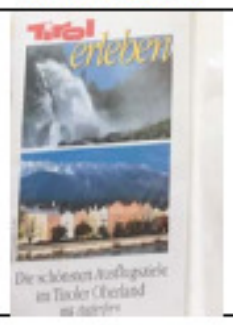 | 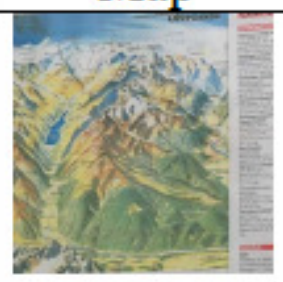 | 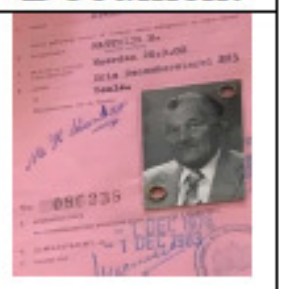 |
| 4 | 8 | 3 | 2 | 1 | 1 | 2 |

### 5.3.2 ACCEPTANCE OF SLOTS-MEMENTO

The elderly agreed that vintage style with decorative effect was unobtrusive when put it at home. One elderly interviewee said: *"I already have too many household appliances in my home, I don't want one more. I need a vintage product with decoration effects."* Most elderly interviewees showed great interests in Slots-Memento, especially the intuitive operation of the handle. To their opinions, sharing memento stories and the handle operation were the most interesting function. The young interviewees also showed great interests in listening stories of their family mementos. One concern of the elderly is that they worried that nobody cares about their stories of mementos.

We provided the feedback from the young to the elderly: the young generation actually were interested in stories of their family mementos.

### 5.3.3 INTERACTING WITH SLOTS-MEMENTO

In the respect of interaction design, as most interviewees felt difficulty in using digital devices, the metaphor of slots machine in our prototype was understood and accepted by them. Although the elderly were usually afraid of using digital devices, for this prototype, they felt not difficulties to use. Most of the interviewees preferred to use the prototype together with their family members, rather than over a distance.

### 5.3.4 SUGGESTIONS FOR IMPROVEMENT

The portability of Slots-Memento needs to be considered as the elderly would like to use it whenever and wherever possible. Although there are only two operation components on the current prototype: a lever and a big button, the elderly still need concrete operation procedure. Some elderly considered the word "story" was too formal to them, they preferred "past experiences". In the respect of interaction design, one interviewee suggested that a timer or indicator light should be added to provide immediate feedback when the REC/STOP button was pressed. The other interviewee suggested that the sensitivity of operation should be reduced.

### 5.4 CONCLUSION

Slots-Memento was understood and accepted by the elderly users, especially its combination of metaphor of slots machine and tangible interface. The photos of memento were easy to recall memories. It enabled the elderly people to be aware of the stories behind the family mementos, as well as aroused their desire to share them with family members.

## 6. DISCUSSION AND FUTURE WORK

Based on the view that the elderly people are considered as digital content producers, Slots-Memento, a system utilized with the metaphor of Slots machine, aiming to facilitate inter-generational memento story sharing and preservation of the elderly. Stories told by the elderly not only need to be triggered and digitalized, but also delivered. Therefore, two significant considerations should be emphasized: storytelling process and sharing process.

### 6.1 TANGIBLE INTERFACE EMPLOYING METAPHOR REDUCES USING BARRIERS

The Slots-Memento device explores how to integrate familiar and existing operation into a novel device. Our solution is adopting the metaphor of slot machine, which provides intuitive and familiar interaction. Moreover, the unknown next picture arouses users' expectations and curiosities to some extent when they pull down the knob.

From the perspective of design, Slots-Memento device is an interactive device with tangible interface. The classic aesthetic of the Slots-Memento make it unobtrusive when put it at home, which would encourage and attract the elderly user to use it. Slots-Memento not only makes the photos and stories of mementos accessible and visible, it also serves as a tangible reminder for the elderly of what it holds.

### 6.2 FACILITATE MEMENTO STORY SHARE IN A SUSTAINABLE WAY

Slots-Memento is not a single product used by the elderly, but a system containing three components. It is difficult to drive the elderly to use the prototype continuously if there is only one product. In the process of story sharing, triggered by the mementos, the elderly tell stories and which are then conveyed to the young, the young provide feedback to the elderly, and the feedback may also act as memory cue. In this case, the story sharing circulation is sustainable.

Figure 8 shows the system architecture of Slots-Memento. The young are memento providers that copy photos of family mementos to flash drive. Triggered by the mementos, the storyteller (in our case is the elderly) tells stories and which are saved in flash disk. Flash disk is then conveyed to the listeners (in our case are the family members). Listeners could listen and keep the stories. Listeners may also provide feedback to the storyteller after listening to the stories,

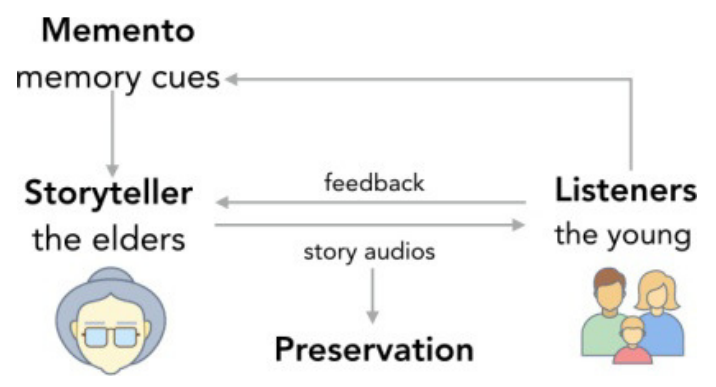

Figure8. System architecture of Slots-Memento

## 6.3 Intergenerational Collaboration Brings Intergenerational Communication

On the one hand, mementos are effective memory triggers, on the other hand, there are lots of stories related to the mementos. Stories related to mementos are right there, but we are unware of it and they need to be excavated. The Slots-Memento system includes two steps, memento digitalization and memory archive. As the elderly are not skilled at technology, but have rich knowledge of family mementos. The whole process is achieved by intergenerational collaboration, which brings intergenerational communication and conversation.

## 6.4 Long-Term Experiment In Future Work

The prototype will be modified and improved based on the feedback of the evaluation part. A long term experiment will be conducted afterwards, prototypes will be distributed to participants and used in real scenarios. Subsequently, semi-structured interviews will be conducted with the elderly people and their family members after they experienced Slots-Memento. Finally, stories told, and mementos collected by the elderly will be analysed.